\documentclass[pre,twocolumn,aps,amsmath,showpacs,amsfonts,amssymb]{revtex4}
\usepackage{graphicx}
\usepackage{times}
\usepackage[utf8]{inputenc}
\usepackage[spanish]{babel}
\usepackage{fancyhdr}

\newcommand{\abs}[1]{\vert #1\vert}
\newcommand{\ecref}[1]{Eq.~\eqref{#1}}

\renewcommand{\theequation}{\thesection.\arabic{equation}}
\begin{document}

\title{Pairwise Summation Approximation of Casimir energy from first principles}
\author{P. Rodriguez-Lopez}
\affiliation{Departamento de F\'\i sica Aplicada I and GISC,
Facultad de Ciencias F\'\i sicas, Universidad Complutense. 28040 Madrid, Spain.}

\begin{abstract}
We obtain the Pairwise Summation Approximation (PSA) of the Casimir energy from first principles in the soft dielectric and soft diamagnetic limit, this analysis let us find that the PSA is an asymptotic approximation of the Casimir energy valid for large distances between the objects. We also obtain the PSA for the electromagnetic (EM) coupling part of the Casimir energy, so we are able to complete the PSA limit for the first time for the complete electromagnetic field.
\end{abstract}


\maketitle

%42.50.Lc 	Quantum fluctuations, quantum noise, and quantum jumps 

% Aquí el título
\title{{\Huge Asymptotic electromagnetic Casimir energy}\\
\author{Pablo Rodriguez} }
\maketitle
%\newpage
%\tableofcontents
%\newpage
\setcounter{equation}{0}
\section{Introduction}
Since 1948, when Casimir introduced the energy that got his name \cite{Casimir Placas Paralelas}, calculation formulas have been looked for. Many analytical and numerical methods have been proposed, such as the zeta function technique, the heat kernel method, semiclassical methods or Green function (local) methods just to mention a few of them \cite{Review Casimir}. However, exact results have been obtained only for some simple geometries. 

The asymptotic analysis of Casimir energies has also a long history. In fact, Casimir himself and Polder in year 1948 gave the first asymptotic formula for the Casimir energy between two electrically neutral bodies in terms of their electric induced dipoles \cite{VdW int. electrica}. Some time later, a generalization of that formula, known as the Pairwise Summation Approximation (PSA) was derived \cite{Lifshitz} for electric media. The main assumption is a linear superposition of the Casimir - Polder interactions between the induced polarizabilities of each element of volume body. Then, the PSA energy is expressed as an integral over the two object's volumes and it is proportional to the objects polarizabilities. The formula has been recently reobtained by R. Golestanian in \cite{Golestanian} and by K. A. Milton et. al. in \cite{Milton} in the soft dielectric limit.

Besides, a new asymptotic method for calculating Casimir energies in term of the induced multipoles of the interacting bodies has been proposed in \cite{Kardar-Geometrias-Arbitrarias} and \cite{Functional Determinant Method 2}. This formula provides a procedure for the calculation of the Casimir energy between \textit{N} arbitrary shaped compact objects \cite{Rodriguez-Lopez 1}.

Our goal is the asymptotic calculation of the electromagnetic Casimir energy and a presentation of a systematic asymptotic expansion procedure for the integral dipole formula on higher orders. We will reobtain the classical results of Casimir and Polder \cite{VdW int. electrica}, and of Feinberg-Sucher in \cite{VdW int. magnetica} in the rarified or soft dielectric and soft diamagnetic limit.

The result we obtain here is a generalization of Milton et. al. one \cite{Milton} but assuming that the bodies are soft diamagnetic as well as soft dielectric. For this purpose, we will use the multi-scattering expansion of the Casimir energy formalism given in \cite{Kardar-Geometrias-Arbitrarias}. To our knownledge, it is the first time this formalism is used to derive the PSA. We obtain that the PSA is the first order of a perturbation expansion in the difference between the electric and magnetic permeability constants of the objects respect the electric and magnetic permeability constants of the medium were the objects are placed. The interest of this result is that we obtain this pairwise summation formula for the complete electromagnetic Casimir energy which is derived as an asymptotic limit of an exact and free of divergences formula. That means that now we can establish the range of validity for that approximation. In fact, we will justify why this formula is valid in the far distance objects limit. This derivation also gives us a perturbative procedure for corrections of this approximation and  posterior expansions to more than two objects or finite temperature cases.

We will follow this plan for the article:
Using the soft dielectric and soft diamagnetic approximation, in Sect. \ref{sec: 2} we will obtain the PSA of Casimir energy for the zero temperature case starting from the exact Casimir energy formula given in \cite{Kardar-Geometrias-Arbitrarias}. We will also study the far distance limit to reobtain the asymptotic Casimir energies given in \cite{VdW int. electrica} and \cite{VdW int. magnetica}. In Sect. \ref{sec: 3} we will obtain PSA Casimir energy formulas for any temperature, studying the high and low temperature limits. In Sect. \ref{sec: 4} we will study the PSA Casimir energy for the three bodies system and for the general $N$ bodies system. We will obtain a kind of superposition principle of Casimir energy in the PSA approximation. Finally, in Sect. \ref{sec: 5} we will study the first perturbation energy term to the PSA approximation and we will discuss about the nature of the PSA limit of the Casimir energy.

\setcounter{equation}{0}
\section{Diluted limit at zero temperature}\label{sec: 2}
Our objective is the calculation of the complete electromagnetic Casimir energy between two bodies in the soft dielectric limit. For this purpose, we will use the Casimir energy formula between two compact bodies at temperature $T = 0$, given in \cite{Kardar-Geometrias-Arbitrarias} as
\begin{equation}\label{Formula de Emig}
E = \frac{\hbar c}{2\pi}\int_{0}^{\infty}dk\log\left(\abs{\mathbb{I} - \mathbb{N}}\right).
\end{equation}
Where $k$ is the frequency, $\mathbb{I}$ is the identity matrix, $\mathbb{N}$ is the matrix $\mathbb{N} = \mathbb{T}_{1}\mathbb{U}_{12}\mathbb{T}_{2}\mathbb{U}_{21}$. Here, $\mathbb{T}_{\alpha}$ is the T scattering matrix of the $\alpha$ - body under the electromagnetic field, and $\mathbb{U}_{\alpha\beta}$ is the propagation matrix of the electromagnetic field from $\alpha$ - body to $\beta$ - body. We will use the position representation instead the multipole representation used in \cite{Kardar-Geometrias-Arbitrarias}, so we can identify $\mathbb{U}_{\alpha\beta} = G_{0\alpha\beta}$, where $G_{0\alpha\beta}$ is the free dyadic Green function. Taking into account that $\log(\abs{A}) =\textrm{Tr}(\log(A))$ and that $\log(1 - x) = - \sum_{p=1}^{\infty}\frac{x^{p}}{p}$, we transform \ecref{Formula de Emig} into
\begin{equation}\label{Formula de Emig en forma de traza}
E = - \frac{\hbar c}{2\pi}\sum_{p=1}^{\infty}\frac{1}{p}\int_{0}^{\infty}dk\textrm{Tr}\left(\mathbb{N}^{p}\right).
\end{equation}
Equation \eqref{Formula de Emig en forma de traza} is an asymptotic expansion of \ecref{Formula de Emig} in the $p_{max}<\infty$ case, which means that our calculus will be valid in the large bodies distance limit. The T operator is related with the potential $V$ (we will see what $V$ is later) by the Lippmann - Schwinger equation, that can be written as \cite{Galindo y Pascual}
\begin{equation}\label{Lippmann - Schwinger equation for T operator}
\mathbb{T} = \left(1 - VG_{0}\right)^{-1}V.
\end{equation}
Applying a Born expansion to \ecref{Lippmann - Schwinger equation for T operator}, we can obtain an approximation of the $T$ operator in the soft dielectric and soft diamagnetic limit as
\begin{equation}\label{Born expansion of T operator}
\mathbb{T} = \sum_{n=0}^{\infty}(VG_{0})^{n}V\simeq V.
\end{equation}
This approximation is more valid for weaker $V$, so here is where the soft dielectric and soft diamagnetic limit is applied.

Now we will study the lowest expansion order. In the lowest expansion order ($p=1$ and $\mathbb{T}=V$) we get the asymptotic approximation of the Casimir energy between two bodies as
\begin{equation}\label{Formula de Emig en soft limit}
E = - \frac{\hbar c}{2\pi}\int_{0}^{\infty}dk\textrm{Tr}\left(V_{1}G_{012}V_{2}G_{021}\right).
\end{equation}
We can separate the magnetic and electric part of each body potential (which is a diagonal operator in positions space, because it is local) and each field contribution in the free dyadic Green function. For isotropic dielectrics with constant electric and magnetic permeabilities, it is easy to find that $V_{\alpha}^{E}(\textbf{r}) = (\epsilon_{\alpha} - \epsilon_{0})\chi_{\alpha}(\textbf{r}) = \tilde{\epsilon}_{\alpha}\chi_{\alpha}(\textbf{r})$ and $V_{\alpha}^{H}(\textbf{r}) = (\mu_{\alpha} - \mu_{0})\chi_{\alpha}(\textbf{r}) = \tilde{\mu}_{\alpha}\chi_{\alpha}(\textbf{r})$ in the vacuum, where $\chi_{\alpha}(\textbf{r})$ is the characteristic function of the $\alpha$ - body volume (1 inside the body and 0 in the rest of the space). Note that we represent the electromagnetic properties of the objects by their potential energy instead the complement representation by their boundary conditions \cite{Balian and Duplantier II}. It will let us to obtain the PSA as an integral over the volume of the bodies as required by the PSA. The contributions of the free dyadic Green function are obtained in the Appendix in \ecref{Acoplo EE funcion Green} - \eqref{Acoplo HH funcion Green}.

As seen in \ecref{Acoplo EE funcion Green} - \eqref{Acoplo HH funcion Green}  of the Appendix, operators are defined over three different linear spaces:
1) An $EH$ - space, whose components are the electric and the magnetic field; 2) over the space coordinates, because we are working with a vector and a pseudovector; and 3) over positions. We must solve the trace over these three spaces: $EH$ - space, vector coordinate space and position space. First we solve the trace in the $EH$ - space and we obtain:
\begin{eqnarray}
E & = & - \frac{\hbar c}{2\pi}\int_{0}^{\infty}dkTr(V_{1}^{E}G_{012}^{EE}V_{2}^{E}G_{021}^{EE} \nonumber\\
  &   &  + V_{1}^{E}G_{012}^{EH}V_{2}^{H}G_{021}^{HE} + V_{1}^{H}G_{012}^{HE}V_{2}^{E}G_{021}^{EH} \nonumber\\
  &   &  + V_{1}^{H}G_{012}^{HH}V_{2}^{H}G_{021}^{HH}).
\end{eqnarray}
We identify each term with this obvious notation:
\begin{equation}\label{Energia de Milton}
E = E_{EE} + E_{EH} + E_{HE} + E_{HH}.
\end{equation}
\subsection{Purely electric and purely magnetic Casimir energy}
The purely electric case has been solved by K. A. Milton et. al. in \cite{Derivacion caso ee por Milton}, where they give the following result:
\begin{equation}
E_{EE} = -23\tilde{\epsilon}_{1}\tilde{\epsilon}_{2}\frac{\hbar c}{(4\pi)^{3}}\int_{1}\int_{2}\frac{d\textbf{r}_{1}d\textbf{r}_{2}}{\abs{\textbf{r}_{1} - \textbf{r}_{2}}^{7}}.
\end{equation}
We obtain the same result because the dyadic Green function used in \cite{Derivacion caso ee por Milton} is the pure electric part of the Green function matrix we use here. In fact, the pure magnetic part of the Green function matrix is equal to the pure electric part $G_{0ij}^{HH}(R,k) = G_{0ij}^{EE}(R,k)$, so the calculus of the purely magnetic part of the Casimir energy is similar to the electric one. Consequently we obtain the following result for the purely magnetic part of the Casimir energy:
\begin{equation}
E_{HH} = -23\tilde{\mu}_{1}\tilde{\mu}_{2}\frac{\hbar c}{(4\pi)^{3}}\int_{1}\int_{2}\frac{d\textbf{r}_{1}d\textbf{r}_{2}}{\abs{\textbf{r}_{1} - \textbf{r}_{2}}^{7}}.
\end{equation}
To our knowledge, this is the first place where this result is obtained.
\subsection{Coupled electromagnetic Casimir energy}
Here we are going to calculate the contribution of the electromagnetic coupling part of the Casimir energy. Using \ecref{Acoplo EH funcion Green} and \ecref{Energia de Milton}, we have to solve:
\begin{equation}
E_{EH} = - \frac{\hbar c}{2\pi}\int_{0}^{\infty}dk\textrm{Tr}\left(V_{1}^{E}G_{012}^{EH}V_{2}^{H}G_{021}^{HE}\right).
\end{equation}
Replacing each potential by its value and assuming isotropy, it is followed that they are proportional to the identity matrix. Then we can drop them for the coordinates trace, but not for the spatial positions trace:
\begin{eqnarray}
E_{EH} & = & - \tilde{\epsilon}_{1}\tilde{\mu}_{2}\frac{\hbar c}{2\pi}\int_{0}^{\infty}dk\int d\textbf{r}_{1}\int d\textbf{r}_{2}\chi_{1}\chi_{2}\nonumber\\
& & \times \textrm{Tr}\left(G_{012}^{EH}G_{021}^{HE}\right).
\end{eqnarray}
Using \ecref{Acoplo EH funcion Green} of the Appendix and $R = \abs{\textbf{r} - \textbf{r}'}$ we have, in matrix form:
\begin{eqnarray}
G_{0ij}^{EH}(R,k) & = & - k\frac{\partial G_{0}}{\partial R}\epsilon_{ijk}\partial_{k}R \nonumber\\
                  & = & G_{0}(R,k)\frac{k}{R}\left(k + \frac{1}{R}\right)\nonumber\\
                  &   & \times\left( \begin{array}{ccc}
0 & R_{3} & - R_{2}\\
- R_{3} & 0 & R_{1}\\
R_{2} & - R_{1} & 0
\end{array} \right),
\end{eqnarray}
so the trace in spatial coordinates is easily solved:
\begin{equation}\label{Traza terminos de acoplo em}
\textrm{Tr}\left(G_{012}^{EH}G_{021}^{HE}\right) = - 2k^{2}G_{0}^{2}(R,k)\left(k + \frac{1}{R}\right)^{2}.
\end{equation}
We can simplify our calculus using the dimensionless variable $u = kR$ to factorize spatial and frequency contributions of the formula
\begin{equation}\label{Energia Casimir acoplo em a T = 0}
E_{EH} = \tilde{\epsilon}_{1}\tilde{\mu}_{2}\frac{4\hbar c}{(4\pi)^{3}}\alpha\int d\textbf{r}_{1}\int d\textbf{r}_{2}\frac{\chi_{1}\chi_{2}}{R^{7}},
\end{equation}
where
\begin{equation}
\alpha = \int_{0}^{\infty}due^{-2u}\left(u^{4} + 2u^{3} + u^{2}\right).
\end{equation}
Finally, using $\int_{0}^{\infty}duu^{n}e^{-au} = \frac{n!}{a^{n+1}}$, the final result is obtained as
\begin{equation}
E_{EH} = 7\tilde{\epsilon}_{1}\tilde{\mu}_{2}\frac{\hbar c}{(4\pi)^{3}}\int d\textbf{r}_{1}\int d\textbf{r}_{2}\frac{\chi_{1}\chi_{2}}{R^{7}}.
\end{equation}
There is an antisymmetry between $G^{EH}$ and $G^{HE}$ shown in \ecref{Acoplo EH funcion Green} and \ecref{Acoplo HE funcion Green} of the Appendix, that is $G_{0ij}^{EH}(R,k) = - G_{0ij}^{HE}(R,k)$. Therefore, we can obtain the coupling between the magnetic part of the first object and the electric part of the second one in a similar way.
Then, we can obtain the coupling between the magnetic part of the first object and the electric part of the second one:
\begin{equation}
E_{HE} = 7\tilde{\epsilon}_{2}\tilde{\mu}_{1}\frac{\hbar c}{(4\pi)^{3}}\int_{1}\int_{2}\frac{d\textbf{r}_{1}d\textbf{r}_{2}}{\abs{\textbf{r}_{1} - \textbf{r}_{2}}^{7}}.
\end{equation}
These two cross contributions to the Casimir energy in the soft dielectric and diamagnetic limits are the main result of this article. It is interesting to note the sign change in this part of the Casimir energy with respect to the other contributions. Then, these terms can invert the typical attractive nature of the Casimir energy to repulsive in this limit for objects with very different electromagnetic nature. Finally, thanks to these new two terms, we obtain the complete electromagnetic Casimir energy between two objects in the soft dielectric and diamagnetic limit for the first time.
Then the global asymptotic electromagnetic Casimir energy is finally obtained:
\begin{equation}\label{Final Result}
E = \frac{- \hbar c}{(4\pi)^{3}} \gamma \int_{1}\int_{2}\frac{d\textbf{r}_{1}d\textbf{r}_{2}}{\abs{\textbf{r}_{1} - \textbf{r}_{2}}^{7}},
\end{equation}
where $\gamma = 23\tilde{\epsilon}_{1}\tilde{\epsilon}_{2} - 7\tilde{\epsilon}_{1}\tilde{\mu}_{2} - 7\tilde{\epsilon}_{2}\tilde{\mu}_{1} + 23\tilde{\mu}_{1}\tilde{\mu}_{2}$ is the complete multiplicative constant, taking into account all the electromagnetic effects in this limit, nor just electric effects, where $\gamma$ would be $23\tilde{\epsilon}_{1}\tilde{\epsilon}_{2}$.
\subsection{Asymptotic Casimir energy}\label{sec: 3.C}
We can also obtain from \ecref{Final Result} (always in the soft electric and magnetic limit) the first asymptotic energy order for two objects. These formulas are the first order of a multipolar expansion of the integrand in \ecref{Final Result} in the coordinate system of each object. That means that we have to assume that the distance between the objects is much greater than their characteristic lengths $R_{\alpha}$, that is $R_{\alpha}\ll R$. In this limit we can separate the problem into two scales and we can replace $\abs{\textbf{r}_{1} - \textbf{r}_{2}} = R$, where $R$ is assume to be a constant. Then the integral can be easily solved in this limit to:
\begin{equation}
E \simeq \frac{- \hbar c}{(4\pi)^{3}}\gamma\frac{V_{1}V_{2}}{R^{7}},
\end{equation}
where $V_{\alpha}$ is the volume of the $\alpha$ - object. In the soft limit order we can approximate the electric and magnetic polarizabilities as $\alpha^{E} = \tilde{\epsilon}\frac{V}{4\pi}$ and $\alpha^{H} = \tilde{\mu}\frac{V}{4\pi}$. In \cite{Optica de Wolf} it was derived $\alpha^{E}$ for an sphere. By using the same method with the approximation of that the effective field over the dielectric is equal to the induced field in the soft limit, we arrive at $\alpha^{E} = \tilde{\epsilon}\frac{V}{4\pi}$ for any arbitrary shaped object. Equation \eqref{Final Result} simplifies into
\begin{small} \begin{equation}
E \simeq\frac{- \hbar c}{4\pi R^{7}}\left(23\alpha_{1}^{E}\alpha_{2}^{E} + 23\alpha_{1}^{H}\alpha_{2}^{H} - 7\alpha_{1}^{E}\alpha_{2}^{H} - 7\alpha_{1}^{H}\alpha_{2}^{E}\right).
\end{equation}\end{small}
This is the soft response limit of the Feinberg and Sucher potential \cite{VdW int. magnetica}. It coincide as well with the soft response limit of the asymptotic Casimir energy obtained in \cite{Kardar-Geometrias-Arbitrarias} for spheres. The presented scheme looks as we would use a local two points potential (where we substitute the polarizabilities of the bodies by their local susceptibilities) as in the present case with the potential given in \ecref{Final Result}, and integrate over the volume of each body to obtain the PSA of the Casimir energy. If we also study the asymptotic distance limit of the PSA, then we reobtain the Feinberg and Sucher potential in the soft response limit, now proportional to the polarizabilities of the objects instead the susceptibilities.

\setcounter{equation}{0}
\section{Diluted limit at any temperature $T$}\label{sec: 3}
In this section we will calculate the Casimir energy in the diluted limit at any finite temperature. Then we will focus on different approximations to low an high temperature limits, recovering the zero temperature case showed before and giving new formulas of the PSA for any temperature. We begin this study with the Casimir energy formula \eqref{Formula de Emig} for any temperature:
\begin{equation}\label{Formula de Emig T}
E = k_{B}T{\sum_{n = 0}^{\infty}}'\log\left(\abs{\mathbb{I} - \mathbb{N}(k_{n})}\right),
\end{equation}
with Matsubara frequencies $k_{n} = 2\pi\frac{k_{B}T}{\hbar c}n = \Lambda n$. The tilde means that the $n=0$ case is weighted by a $1/2$ factor. As usual, we apply the Born approximation to the $T$ matrix scattering obtaining at first order $\mathbb{T}_{\alpha} = V_{\alpha}$ (where $\alpha$ labels the body in interaction). We apply again that $\log(\abs{A}) = Tr(\log(A))$ and $\log(1 - x) = - \sum_{p=1}^{\infty}\frac{x^{p}}{p}$ , then we transform \ecref{Formula de Emig T} into:
\begin{equation}
E = - k_{B}T\sum_{p=1}^{\infty}\frac{1}{p}{\sum_{n = 0}^{\infty}}'\textrm{Tr}\left(\mathbb{N}^{p}(\Lambda n)\right).
\end{equation}
At first order in $p$ we obtain:
\begin{eqnarray}
E & = & - k_{B}T{\sum_{n = 0}^{\infty}}'\textrm{Tr}\left(\mathbb{N}(\Lambda n)\right)\nonumber\\
  & = & - k_{B}T{\sum_{n = 0}^{\infty}}'\textrm{Tr}\left(\mathbb{T}_{1}\mathbb{U}_{12}\mathbb{T}_{2}\mathbb{U}_{21}\right).
\end{eqnarray}
As we are working in the positions representation space instead in the multipolar representation space, we have to represents the operators $\mathbb{T}_{\alpha}$ and $\mathbb{U}_{\alpha\beta}$ in the positions space. For that issue, we have into account that $\mathbb{U}_{\alpha\beta}$ matrices represent the field propagation between two points. Then they are represented by the free vacuum Green function of the interaction field. Using the Born approximation, we obtain the next formula for the diluted approximation of the Casimir energy:
\begin{equation}
E \simeq E_{T} = - k_{B}T{\sum_{n = 0}^{\infty}}'\textrm{Tr}\left(V_{1}G_{012}V_{2}G_{021}\right).
\end{equation}
As we did with \eqref{Formula de Emig en soft limit}, we can separate the magnetic and electric part of each body potential and each field contribution in the free dyadic Green function. We trace over the $EH$ space obtaining:
\begin{eqnarray}
E_{T} & = & - k_{B}T{\sum_{n = 0}^{\infty}}'Tr(V_{1}^{E}G_{012}^{EE}V_{2}^{E}G_{021}^{EE} + \nonumber\\
  &   & + V_{1}^{E}G_{012}^{EH}V_{2}^{H}G_{021}^{HE} + V_{1}^{H}G_{012}^{HE}V_{2}^{E}G_{021}^{EH}\nonumber\\
  &   & + V_{1}^{H}G_{012}^{HH}V_{2}^{H}G_{021}^{HH}).
\end{eqnarray}
As before, we identify each term with this obvious notation:
\begin{equation}
E_{T} = E_{EE} + E_{EH} + E_{HE} + E_{HH}.
\end{equation}
\subsection{Purely electric and purely magnetic energy}
Using \ecref{Acoplo EE funcion Green} and \ecref{Acoplo HH funcion Green} of the Appendix, we can solve the trace over the coordinates of $E_{EE}$ and, similarly, of $E_{HH}$. The matricial form of the purely electric part of the dyadic Green function \label{Acoplo EE funcion Green} and of the purely magnetic part of the dyadic Green function\label{Acoplo HH funcion Green} are:
\begin{eqnarray}
G_{0ij}^{EE}(R,k) & = & - \left( \begin{array}{ccc}
R_{x}^{2} & R_{x}R_{y} & R_{x}R_{z}\\
R_{y}R_{x} & R_{y}^{2} & R_{y}R_{z}\\
R_{z}R_{x} & R_{z}R_{y} & R_{z}^{2}
\end{array} \right)\nonumber\\
& & \times\left(3 + 3kR + k^{2}R^{2}\right)\frac{e^{-kR}}{4\pi R^{5}}\nonumber\\
      &   & + \left( \begin{array}{ccc}
1 & 0 & 0\\
0 & 1 & 0\\
0 & 0 & 1
\end{array} \right)\nonumber\\
& & \times\left(1 + kR + k^{2}R^{2}\right)\frac{e^{-kR}}{4\pi R^{3}}\nonumber\\
& = & G_{0ij}^{HH}(R,k).
\end{eqnarray}
Where $\textbf{R} = \textbf{r}_{\alpha} - \textbf{r}_{\beta}$. The trace in spatial coordinates is easily solved obtaining:
\begin{eqnarray}
E_{EE} & = & - k_{B}T{\sum_{n = 0}^{\infty}}'\tilde{\epsilon}_{1}\tilde{\epsilon}_{2}\int_{1}d\textbf{r}_{1}\int_{2}d\textbf{r}_{2}\frac{e^{-2kR}}{(4\pi)^{2} R^{6}}\nonumber\\
& & \times[6 + 12kR + 10k^{2}R^{2} + 4k^{3}R^{3} + 2k^{4}R^{4}].\nonumber\\
& & \label{Caso diagonal T neq 0}
\end{eqnarray}
Replacing $k$ by $\Lambda n$ we obtain:
\begin{eqnarray}
E_{EE} & = & - \frac{k_{B}T}{(4\pi)^{2}}\tilde{\epsilon}_{1}\tilde{\epsilon}_{2}\int_{1}\int_{2}\frac{d\textbf{r}_{1}d\textbf{r}_{2}}{R^{6}}{\sum_{n = 0}^{\infty}}'e^{-2\Lambda Rn}\nonumber\\
& & \times[6 + 12\Lambda Rn + 10\Lambda^{2}R^{2}n^{2}\nonumber\\
& & + 4\Lambda^{3}R^{3}n^{3} + 2\Lambda^{4}R^{4}n^{4} ]\label{Caso diagonal T neq 0 2}.
\end{eqnarray}
Having into account that $\sum_{n = 0}^{\infty}e^{-an} = \frac{1}{1 - e^{-a}}$, it is easily deduced:
\begin{equation}
\partial_{a}^{t}\sum_{n = 0}^{\infty}e^{-an} = \sum_{n = 0}^{\infty}(-n)^{t}e^{-an} = \partial_{a}^{t}(1 - e^{-a})^{-1}.
\end{equation}
Denoting $\lambda = R\Lambda$, we can carry out the sum obtaining:
\begin{eqnarray}\label{Caso diagonal T neq 0 resuelto}
E_{EE} & = & - \frac{k_{B}T}{(4\pi)^{2}}\tilde{\epsilon}_{1}\tilde{\epsilon}_{2}\int_{1}\int_{2}\frac{d\textbf{r}_{1}d\textbf{r}_{2}}{R^{6}}\frac{1}{(e^{2\lambda} - 1)^{5}}\nonumber\\
& & \times[e^{2\lambda}\left( - 9 - 12\lambda + 10\lambda^2 - 4\lambda^3 + 2\lambda^4 \right)\nonumber\\
& & + e^{4\lambda}\left(  6 + 36\lambda - 10\lambda^2 - 12\lambda^3 + 22\lambda^4 \right)\nonumber\\
& & + e^{6\lambda}\left(  6 - 36\lambda - 10\lambda^2 + 12\lambda^3 + 22\lambda^4 \right)\nonumber\\
& & + e^{8\lambda}\left(- 9 + 12\lambda + 10\lambda^2 +  4\lambda^3 +  2\lambda^4 \right)\nonumber\\
& & + 3 e^{10\lambda} + 3].
\end{eqnarray}
We obtain a similar result for the purely magnetic Casimir energy replacing $\tilde{\epsilon}_{1}\tilde{\epsilon}_{2}$ by $\tilde{\mu}_{1}\tilde{\mu}_{2}$.
\subsection{Coupled magnetic - electric Casimir energy terms}
We perform the same calculations using \eqref{Acoplo EH funcion Green} and \eqref{Acoplo HE funcion Green} of the Appendix as for the purely electric case. The trace over spatial coordinates is already done in \eqref{Traza terminos de acoplo em}, so we obtain the formula for the coupling terms of the Casimir energy just replacing the integral in \eqref{Energia Casimir acoplo em a T = 0} by a sum:
\begin{eqnarray}
E_{EH} & = & \frac{2k_{B}T}{(4\pi)^{2}}\tilde{\epsilon}_{1}\tilde{\mu}_{2}\int_{1}\int_{2}\frac{d\textbf{r}_{1}d\textbf{r}_{2}}{R^{6}}{\sum_{n = 0}^{\infty}}'e^{-2kR}\nonumber\\
& & \times\left(R^{4}k^{4} + 2R^{3}k^{3} + R^{2}k^{2}\right)\label{Energia Casimir acoplo em a T neq 0}.
\end{eqnarray}
Here $k_{n} = 2\pi\frac{k_{B}T}{\hbar c}n = \Lambda n$ are the Matsubara frequencies. After solving this sum and denoting $\lambda = R\Lambda$, we obtain:
\begin{eqnarray}\label{Energia Casimir acoplo em a T neq 0 resuelto}
E_{EH} & = & \frac{2k_{B}T}{(4\pi)^{2}}\tilde{\epsilon}_{1}\tilde{\mu}_{2}\int_{1}\int_{2}\frac{d\textbf{r}_{1}d\textbf{r}_{2}}{R^{6}}\frac{\lambda^2}{(e^{2\lambda} - 1)^{5}}\nonumber\\
& & \times[e^{2\lambda}\left(  1 - 2\lambda +\lambda^2 \right)\nonumber\\
& & + e^{4\lambda}\left( -1 - 6\lambda + 11\lambda^2\right)\nonumber\\
& & + e^{6\lambda}\left( -1 + 6\lambda + 11\lambda^2\right)\nonumber\\
& & + e^{8\lambda}\left( 1 + 2\lambda +\lambda^2 \right) ].
\end{eqnarray}
Replacing $\tilde{\epsilon}_{1}\tilde{\mu}_{2}$ by $\tilde{\mu}_{1}\tilde{\epsilon}_{2}$, we obtain the formula for $E_{HE}$. These formulas for the Casimir energy in the diluted limit are valid for any temperature, but they are too much complicate for analytical analysis at any temperature. Therefore, we study the limits at high and low temperatures.
\subsection{Low and zero temperature limit}
Here we will recover the zero temperature limit of the diluted limit. Then we will make a perturbative analysis valid for low temperatures. We can make it easily by expanding the Taylor series in $\lambda = R\Lambda$ of the integrand and studying just the first orders deleting the rest ones. By using \eqref{Caso diagonal T neq 0 resuelto} and \eqref{Energia Casimir acoplo em a T neq 0 resuelto},  we reobtain \eqref{Final Result} for the zero temperature case. We need to take the fifth order series term of the Taylor expansion of \ecref{Caso diagonal T neq 0 resuelto} and \ecref{Energia Casimir acoplo em a T neq 0 resuelto} to get the next non zero perturbation term of the Casimir energy as
\begin{equation}
\frac{\Delta_{5}E_{EE}}{\tilde{\epsilon}_{1}\tilde{\epsilon}_{2}} = \frac{\Delta_{5}E_{HH}}{\tilde{\mu}_{1}\tilde{\mu}_{2}} = - \frac{22\pi^{3}}{945}k_{B}T\left(\frac{k_{B}T}{\hbar c}\right)^{5}\int_{1}\int_{2}\frac{d\textbf{r}_{1}d\textbf{r}_{2}}{R},
\end{equation}
\begin{equation}
\frac{\Delta_{5}E_{EH}}{\tilde{\epsilon}_{1}\tilde{\mu}_{2}} = \frac{\Delta_{5}E_{HE}}{\tilde{\mu}_{1}\tilde{\epsilon}_{2}} = - \frac{2\pi^{3}}{189}k_{B}T \left(\frac{k_{B}T}{\hbar c}\right)^{5}\int_{1}\int_{2}\frac{d\textbf{r}_{1}d\textbf{r}_{2}}{R},
\end{equation}
because $\Delta_{1}E = \Delta_{2}E = \Delta_{3}E = \Delta_{4}E = 0 $, where $\Delta_{n}E$ is the n-th order correction term to the low temperature expansion.
\subsection{High temperature and classical limit}
In this section we will obtain the high temperature limit of the Casimir energy in the diluted limit, whose first term will be the classical limit of the Casimir energy. To obtain this limit, instead of solving the sum in \ecref{Caso diagonal T neq 0} and \ecref{Energia Casimir acoplo em a T neq 0}, we will just keep the first term of the sum. Then the classical limit of the Casimir energy in the diluted limit reads the first sum term:
\begin{equation}
\frac{E_{EE}^{cl}}{\tilde{\epsilon}_{1}\tilde{\epsilon}_{2}} = \frac{E_{HH}^{cl}}{\tilde{\mu}_{1}\tilde{\mu}_{2}} = - 3 \frac{k_{B}T}{(4\pi)^{2}}\int_{1}\int_{2}\frac{d\textbf{r}_{1}d\textbf{r}_{2}}{R^{6}},
\end{equation}
\begin{equation}
\frac{E_{EH}^{cl}}{\tilde{\epsilon}_{1}\tilde{\mu}_{2}} = \frac{E_{HE}^{cl}}{\tilde{\mu}_{1}\tilde{\epsilon}_{2}} = 0.
\end{equation}
And the first perturbation to that limit is the next sum term:
\begin{eqnarray}
\frac{\Delta_{1}E_{EE}^{cl}}{\tilde{\epsilon}_{1}\tilde{\epsilon}_{2}} & = & - \frac{k_{B}T}{(4\pi)^{2}}\int_{1}\int_{2}\frac{d\textbf{r}_{1}d\textbf{r}_{2}}{R^{6}}e^{-2\Lambda R}\nonumber\\
& & \times\left(6 + 12\Lambda R + 10\Lambda^{2}R^{2} + 4\Lambda^{3}R^{3} + 2\Lambda^{4}R^{4}\right)\nonumber\\
& = &\frac{\Delta_{1}E_{HH}^{cl}}{\tilde{\mu}_{1}\tilde{\mu}_{2}},
\end{eqnarray}
\begin{eqnarray}
\frac{\Delta_{1}E_{EH}^{cl}}{\tilde{\epsilon}_{1}\tilde{\mu}_{2}} & = & \frac{\Delta_{1}E_{HE}^{cl}}{\tilde{\mu}_{1}\tilde{\epsilon}_{2}} = \frac{2k_{B}T}{(4\pi)^{2}}\int_{1}\int_{2}\frac{d\textbf{r}_{1}d\textbf{r}_{2}}{R^{6}}e^{-2R\Lambda}\nonumber\\
& & \times\left(R^{4}\Lambda^{4} + 2R^{3}\Lambda^{3} + R^{2}\Lambda^{2}\right).
\end{eqnarray}
Where $\Delta_{1}E^{cl}$ is the first correction to the classic limit of the Casimir energy $E^{cl}$. With these results we see that we loose the coupling between electric and magnetic energy terms in the classical limit.

Finally we must remark that, if we take the asymptotic distance approximation as in Sect. \ref{sec: 3.C}, we reobtain the results given in \cite{Barton} in the soft response limit.

\setcounter{equation}{0}
\section{PSA for three bodies system}\label{sec: 4}
In this section we are going to calculate the Casimir energy between three bodies in the Pairwise Summation approximation. In this asymptotic limit we will obtain that the energy of the system will be the addition of the PSA energy of each pair of objects. This linear behavior of the Casimir energy was expected in that approximation although we know that it is in general false. We begin this study from the Casimir Energy formula for three objects given in \cite{Emig Casimir caso escalar}:
\begin{equation}\label{Casimir N objects}
E = \frac{\hbar c}{2\pi}\int_{0}^{\infty}dk\log\left(\frac{\abs{\mathbb{M}}}{\phantom{_{\infty}}\abs{\mathbb{M}}_{\infty}}\right).
\end{equation}
The $M$ matrix (whose coefficients are non commutative matrices) is:
\begin{displaymath}
\mathbb{M} = \left(\begin{array}{c c c}
\phantom{-}\mathbb{T}_{1}^{-1} & - \mathbb{U}_{12}   & - \mathbb{U}_{13}\\
 - \mathbb{U}_{21}  & \phantom{-}\mathbb{T}_{2}^{-1} & - \mathbb{U}_{23}\\
 - \mathbb{U}_{31}  & - \mathbb{U}_{32}   & \phantom{-}\mathbb{T}_{3}^{-1}
\end{array}\right).
\end{displaymath}
So, using the logarithm product rule, the Casimir energy between three bodies is
\begin{eqnarray}\label{Energia Casimir 3 cuerpos}
E_{3} & = & \frac{\hbar c}{2\pi}\int_{0}^{\infty}dk\log\left(\abs{\mathbb{I} - \mathbb{N}_{12}}\right) \nonumber\\
      &   &  + \frac{\hbar c}{2\pi}\int_{0}^{\infty}dk\log\left(\abs{\mathbb{I} - \mathbb{N}_{13}}\right) \nonumber\\
      &   &  + \frac{\hbar c}{2\pi}\int_{0}^{\infty}dk\log\left(\abs{\mathbb{I} - \mathbb{R}}\right).
\end{eqnarray}
Here the $N$ matrix is $\mathbb{N}_{\alpha\beta} = \mathbb{T}_{\alpha}\mathbb{U}_{\alpha\beta}\mathbb{T}_{\beta}\mathbb{U}_{\beta\alpha}$, and:
\begin{eqnarray}
\mathbb{R} & = & \left(\mathbb{I} - \mathbb{N}_{13}\right)^{-1}\left(\mathbb{T}_{3}\mathbb{U}_{32} + \mathbb{N}_{31}\right)\nonumber\\
      &   & \times\left(\mathbb{I} - \mathbb{N}_{12}\right)^{-1}\left(\mathbb{T}_{2}\mathbb{U}_{23} + \mathbb{N}_{21}\right).
\end{eqnarray}
Expanding the inverses we obtain the double series
\begin{eqnarray}
\mathbb{R} & = & \sum_{n_{1}=0}^{\infty}\left(\mathbb{N}_{13}\right)^{n_{1}}\left(\mathbb{T}_{3}\mathbb{U}_{32} + \mathbb{N}_{31}\right)\nonumber\\
      &   & \times\sum_{n_{2}=0}^{\infty}\left(\mathbb{N}_{12}\right)^{n_{2}}\left(\mathbb{T}_{2}\mathbb{U}_{23} + \mathbb{N}_{21}\right).
\end{eqnarray}
The lowest order of the Born series of the $\mathbb{R}$ matrix will come from the first order expansion series making $n_{1} = n_{2} = 0$, that is
\begin{equation}
\mathbb{R} \simeq \left(\mathbb{T}_{3}\mathbb{U}_{32} + \mathbb{N}_{31}\right)\left(\mathbb{T}_{2}\mathbb{U}_{23} + \mathbb{N}_{21}\right).
\end{equation}
In addition to that, we just consider the sum term with the minimum number of $\mathbb{T}$ matrices products, because that will be the lowest order expansion in susceptibilities of the $\mathbb{R}$ matrix. That means that we reduce the highly non linear $\mathbb{R}$ matrix to
\begin{equation}\label{sol limit R matrix}
\mathbb{R} \simeq \left(\mathbb{T}_{3}\mathbb{U}_{32}\mathbb{T}_{2}\mathbb{U}_{23}\right) = \mathbb{N}_{23}.
\end{equation}
Replacing \ecref{sol limit R matrix} in \ecref{Energia Casimir 3 cuerpos}, we obtain the next PSA of the Casimir energy between three objects:
\begin{eqnarray}\label{3-bodies PSA}
E_{3} & = & \frac{\hbar c}{2\pi}\int_{0}^{\infty}dk\log\left(\abs{\mathbb{I} - \mathbb{N}_{12}}\right) \nonumber\\
      &   &  + \frac{\hbar c}{2\pi}\int_{0}^{\infty}dk\log\left(\abs{\mathbb{I} - \mathbb{N}_{13}}\right) \nonumber\\
      &   &  + \frac{\hbar c}{2\pi}\int_{0}^{\infty}dk\log\left(\abs{\mathbb{I} - \mathbb{N}_{23}}\right).
\end{eqnarray}
Therefore we obtain that the PSA approximation of the Casimir energy between three objects is the sum of the PSA energy of each pair of objects. In other words, we rederive a kind of superposition law of energies in the diluted PSA limit as expected, because the usual presentation of the PSA approximation is the assumption that we have a superposition behavior in that asymptotic limit. Here we have justified the validity of that approximation.

If we take the asymptotic distance limit to \ecref{3-bodies PSA}, we will obtain a superposition of two bodies PSA energies in the asymptotic limit. The nonlinearity of the Casimir energy must be given by higher orders expansion terms even for systems with three bodies.
The same superposition behavior is expected for the general $N$ body case, and it will be proven in the next section.

\subsection{PSA for general N bodies system}
In this section we are going to generalize the PSA energy of three bodies given by \ecref{3-bodies PSA}. For this purpose we are going to use an iterative procedure which will give us the new terms to include to the PSA of the Casimir energy of $n-1$ bodies when we include another new object to our system. Let us represent the $\mathbb{M}$ matrix of \ecref{Casimir N objects} as the sum of its diagonal and its non-diagonal parts as \cite{Emig Casimir caso escalar}:
\begin{equation}
\mathbb{M}_{\alpha\beta} = \delta_{\alpha\beta}\mathbb{T}_{\alpha}^{-1} + (\delta_{\alpha\beta} - 1)\mathbb{U}_{\alpha\beta},
\end{equation}
or symbolically as
\begin{equation}
\mathbb{M} = \mathbb{T}^{-1} + \mathbb{U},
\end{equation}
then it is easy to find that the inverse of $\mathbb{M}$ is the next perturbative series
\begin{equation}
\mathbb{M}^{-1} = \mathbb{T}\sum_{n=0}^{\infty}\left(-\mathbb{U}\mathbb{T}\right)^{n}.
\end{equation}
On the other hand, the $\mathbb{M}$ matrix of the $N$ objects system is related with the $\mathbb{M}$ matrix of the $N-1$ objects system by block matrices in the next way
\begin{displaymath}
\mathbb{M}_{N} = \left(\begin{array}{c c c}
\phantom{-}\mathbb{M}_{N-1\phantom{,\gamma}} & - \mathbb{U}_{N-1,\gamma}\\
 - \mathbb{U}_{\gamma,N-1}  & \phantom{-}\mathbb{T}_{N\phantom{-1,\gamma}}^{-1}
\end{array}\right).
\end{displaymath}
Where $\gamma$ index goes from 1 to $N-1$. With this result we will calculate the determinant of $\mathbb{M}_{N}$ obtaining
\begin{equation}
\abs{\mathbb{M}_{N}} = \abs{\mathbb{M}_{N-1}}\abs{\mathbb{T}_{N}}^{-1}\abs{\mathbb{I} - \mathbb{T}_{N}\mathbb{U}_{\gamma ,N-1}\mathbb{M}_{N-1}^{-1}\mathbb{U}_{N-1,\gamma}}.
\end{equation}
In the PSA approximation is valid the substitution $\mathbb{M}^{-1} \simeq \mathbb{T}$, where $\mathbb{T}$ is a diagonal matrix whose $N-1$ diagonal elements are $\mathbb{T}_{\gamma}$ with $\gamma$ index defined as before.
Then, using this approximation we can approximate the $\mathbb{M}_{N}$ determinant as:
\begin{equation}
\abs{\mathbb{M}_{N}} = \abs{\mathbb{M}_{N-1}}\abs{\mathbb{T}_{N}}^{-1}\abs{\mathbb{I} - \mathbb{T}_{N}\mathbb{U}_{\gamma ,N-1}\mathbb{T}\mathbb{U}_{N-1,\gamma}}.
\end{equation}
Where we multiply by blocks the matrix of the last determinant obtaining
\begin{eqnarray}
\abs{\mathbb{M}_{N}} & = & \abs{\mathbb{M}_{N-1}}\abs{\mathbb{T}_{N}}^{-1}\abs{\mathbb{I} - \left(\begin{array}{c c c c}
\mathbb{N}_{1N} & 0 & \dots & 0\\
0 & \mathbb{N}_{2N} & \dots & 0\\
\vdots & \vdots & \ddots & \vdots\\
0 & 0 & \dots & \mathbb{N}_{N-1,N}
\end{array}\right)}.\nonumber\\
& &
\end{eqnarray}
Finally, we obtain the desired result:
\begin{equation}
\abs{\mathbb{M}_{N}} = \abs{\mathbb{M}_{N-1}}\abs{\mathbb{T}_{N}}^{-1}\prod_{\gamma =1}^{N-1}\abs{\mathbb{I} - \mathbb{N}_{\gamma N}}.
\end{equation}
The other needed matrix is
\begin{equation}
\abs{\mathbb{M}_{\infty, N}} = \abs{\mathbb{M}_{\infty ,N-1}}\abs{\mathbb{T}_{N}}^{-1}.
\end{equation}
And with the initial conditions:
\begin{equation}
\abs{\mathbb{M}_{1}} = \abs{\mathbb{M}_{\infty, 1}} = \abs{\mathbb{T}_{1}}^{-1},
\end{equation}
we obtain these determinants in a closed form as
\begin{equation}
\abs{\mathbb{M}_{N}} = \prod_{k = 1}^{N}\abs{\mathbb{T}_{k}}^{-1}\prod_{l = 2}^{N}\prod_{m = 1}^{l-1}\abs{\mathbb{I} - \mathbb{N}_{lm}}
\end{equation}
and
\begin{equation}
\abs{\mathbb{M}_{\infty, N}} = \prod_{k = 1}^{N}\abs{\mathbb{T}_{k}}^{-1}.
\end{equation}
Taking the logarithms of \ecref{Casimir N objects}, the Casimir energy for the N bodies system is approximated as
\begin{equation}\label{N-bodies PSA}
E_{N} = \frac{\hbar c}{2\pi}\int_{0}^{\infty}dk\sum_{l = 2}^{N}\sum_{m = 1}^{l-1}\log\left(\abs{\mathbb{I} - \mathbb{N}_{lm}}\right)
\end{equation}
in the PSA limit. \ecref{N-bodies PSA} is the generalization of \ecref{3-bodies PSA} for the $N$ body case, and it show us that in the PSA limit, we always obtain a superposition principle of the two body PSA Casimir energy despite the fact that Casimir energy is not a nonadittive interaction. We will obtain the non-linear effects in the next orders of the expansion.

This result is qualitatively different of the usual PSA procedure of integrating each point of a $N$ point potential over the volume of each object. The reason is simple, a $N$ point potential is proportional to $N$ polarizabilities \cite{Thiru}, but in the soft limit approximation, the lower allowed term is proportional to two polarizabilities. Therefore we should study the $(N-1)$ expansion term to obtain the first one proportional to $N$ polarizabilities.
So in a consistent calculation of PSA energies for three or more objects, the asymptotic approximation of the PSA energy in the soft material approximation is different of the $N$ point potential function.
In the next section we will find where the contribution of the three point potential is relevant in the diluted limit.

\setcounter{equation}{0}
\section{Second order expansion of diluted limit}\label{sec: 5}
In this section we are going to study the second order expansion of the PSA in the diluted limit. This is a complicated long calculus so, instead the complete electromagnetic case, we will restrict ourselves to the purely electric case. That means that we will ignore the magnetic properties of the bodies. It could be interesting to show the complete study of that series term, but it is a long calculation to show here. However, it possesses theoretical utility, as it shows the lack of the superposition principle in the Casimir energy calculations. The next results have been done with the help of {\tt Mathematica} \cite{Mathematica}.

Again we start from the Casimir energy between two compact objects as given by \ecref{Formula de Emig en forma de traza}. We studied the first order expansion of the energy in Sect. \ref{sec: 2}. Now we are interested in the next order expansion term in polarizabilities, so we maintain the study of the fist term of \ecref{Formula de Emig en forma de traza} taking just the case $p = 1$. In addition to that, we take the second order approximation of the Born series of the $\mathbb{T}$ matrix:
\begin{equation}
\mathbb{T} = \sum_{n=0}^{\infty}(VG_{0})^{n}V\simeq V + VG_{0}V.
\end{equation}
Applying the linearity of the trace and taking our attention just to the second order expansion in polarizabilities (that is, to the third order term in polarizabilities), we obtain the following result for the second order of the diluted limit result:
\begin{eqnarray}
E_{2} & = & - \frac{\hbar c}{2\pi}\int_{0}^{\infty}dk\textrm{Tr}\left(V_{1'}G_{01'1}V_{1}U_{12}V_{2}U_{21'}\right) \nonumber\\
      &   & - \frac{\hbar c}{2\pi}\int_{0}^{\infty}dk\textrm{Tr}\left(V_{1}U_{12'}V_{2'}G_{02'2}V_{2}U_{21}\right)\label{segundo orden PSA}.
\end{eqnarray}
We will center our study just on the first integral of \eqref{segundo orden PSA}, because the analysis of these two integrals is the same. First we make the trace over the $EH$ space, which is automatic here because we have cancelled the magnetic properties of the bodies. In other words, we are not studying the coupling between electric and magnetic induced dipoles. After that we trace over the space coordinates, so we need the matricial form of the electric part of the dyadic Green function given in \ecref{Acoplo EE funcion Green} of the Appendix, where $\textbf{R} = \textbf{r} - \textbf{r}'$. We will need to define the function:
\begin{eqnarray}
\mathcal{R}[\alpha ,\beta]_{i}(k) & = & R_{\alpha\beta}^2 \left(1+k R_{\alpha\beta}+k^2 R_{\alpha\beta}^2\right)\nonumber\\
& & + \left(3+3 k R_{\alpha\beta}+k^2 R_{\alpha\beta}^2\right)R_{i\alpha\beta}^2.
\end{eqnarray}
Where $k$ is the frequency, $\alpha$ and $\beta$ labels the bodies coordinates and $i$ shows the vector component of $\textbf{R}_{\alpha\beta}$ used. So we can solve the trace over the space coordinates obtaining the following result for the first term of \eqref{segundo orden PSA}:
\begin{eqnarray}
E_{2a} & = & - \frac{\hbar c}{2\pi}\int_{1}d\textbf{r}_{1}\int_{2}d\textbf{r}_{2}\int_{1'}d\textbf{r}_{1'}\int_{0}^{\infty}dk\nonumber\\
      &   & \times\frac{e^{-k (R_{12} + R_{21'} + R_{1'1})}}{64\pi^{3}R_{12}^{5}R_{21'}^{5}R_{1'1}^{5}}\tilde{\epsilon}_{1}\tilde{\epsilon}_{2}\tilde{\epsilon}_{1'}\nonumber\\
& & \times[\mathcal{R}[1,2]_{x}(k)\mathcal{R}[2,1']_{x}(k)\mathcal{R}[1',1]_{x}(k)\nonumber\\
& & + \mathcal{R}[1,2]_{y}(k)\mathcal{R}[2,1']_{y}(k)\mathcal{R}[1',1]_{y}(k)\nonumber\\
& & + \mathcal{R}[1,2]_{z}(k)\mathcal{R}[2,1']_{z}(k)\mathcal{R}[1',1]_{z}(k)].
\end{eqnarray}
Here the space integrations are performed just over the body volumes because it is there where dielectric and diamagnetic potentials are defined. It is possible to perform the $k$ integral, because we have an exponential multiplied by a polynomial integrated between zero and infinity. It is not shown here because it contains 216 sum terms.

For any finite temperature we also obtain a long result, but in the classical limit we can at least obtain a tractable formula. This limit consists of keeping only the $k = 0$ mode. In that case we obtain the result
\begin{eqnarray}
E_{2a}^{cl} & = & - \frac{\hbar c}{256\pi^{4}}\int_{1}d\textbf{r}_{1}\int_{2}d\textbf{r}_{2}\int_{1'}d\textbf{r}_{1'}\frac{\tilde{\epsilon}_{1}\tilde{\epsilon}_{2}\tilde{\epsilon}_{1'}}{R_{12}^{5}R_{21'}^{5}R_{1'1}^{5}}\nonumber\\
& & \times[\mathcal{R}[1,2]_{x}(0)\mathcal{R}[2,1']_{x}(0)\mathcal{R}[1',1]_{x}(0)\nonumber\\
& & + \mathcal{R}[1,2]_{y}(0)\mathcal{R}[2,1']_{y}(0)\mathcal{R}[1',1]_{y}(0)\nonumber\\
& & + \mathcal{R}[1,2]_{z}(0)\mathcal{R}[2,1']_{z}(0)\mathcal{R}[1',1]_{z}(0)].
\end{eqnarray}
Where $\mathcal{R}[\alpha ,\beta]_{i}(0) = \left(R_{\alpha\beta}^{2} + 3R_{\alpha\beta i}^{2}\right)$, so we can write:
\begin{eqnarray}\label{Divergent integrand example}
E_{2a}^{cl} & = & - \frac{\hbar c}{256\pi^{4}}\int_{1}d\textbf{r}_{1}\int_{2}d\textbf{r}_{2}\int_{1'}d\textbf{r}_{1'}\frac{\tilde{\epsilon}_{1}\tilde{\epsilon}_{2}\tilde{\epsilon}_{1'}}{R_{12}^{5}R_{21'}^{5}R_{1'1}^{5}}\nonumber\\
& & \times[\left(R_{12}^{2} + 3R_{12 x}^{2}\right)\left(R_{21'}^{2} + 3R_{21' x}^{2}\right)\left(R_{1'1}^{2} + 3R_{1'1 x}^{2}\right)\nonumber\\
& & + \left(R_{12}^{2} + 3R_{12 y}^{2}\right)\left(R_{21'}^{2} + 3R_{21' y}^{2}\right)\left(R_{1'1}^{2} + 3R_{1'1 y}^{2}\right)\nonumber\\
& & + \left(R_{12}^{2} + 3R_{12 z}^{2}\right)\left(R_{21'}^{2} + 3R_{21' z}^{2}\right)\left(R_{1'1}^{2} + 3R_{1'1 z}^{2}\right)].\nonumber\\
& &
\end{eqnarray}
When performing this integral, a term proportional to $R_{1'1}^{-5}$ and another one to $R_{1'1}^{-3}$ appear. These terms can be problematic because $R_{1'1} = 0$ belongs to the integration interval for all the points of the $1$ body. But these singularities can be removed with the appropriate regularization procedure.
Note that if we make $1' = 3$ in \ecref{segundo orden PSA}, this integral would also be one of the contributions of the second term expansion of a three objects system. In that case we have not got any problem in the volumes integration for this term and this expansion term looks quite similar to the local analog of the three points potential given in \cite{Thiru}. In fact, in the three (or more) bodies system that term is the first term that breaks the superposition behavior founded in \ref{sec: 4}.
Replacing the $k^{2}$ term by the operator $-\Delta$ in Eqs. \eqref{Acoplo EE funcion Green} and \eqref{Acoplo HH funcion Green} of the Appendix,  and replacing the $k$ term by the operator $\sqrt{-\Delta}$ in Eqs. \eqref{Acoplo HE funcion Green} and \eqref{Acoplo EH funcion Green} of the Appendix, which is a change valid on shell because $G_{0}(R,k) = \frac{e^{-kR}}{4\pi R}$, we recover the formalism used in \cite{Thiru} in the soft dielectric and soft diamagnetic limit. This scheme let us take into account not only electric phenomena as in \cite{Thiru}, but also magnetic and electro-magnetic coupling effects. Taking this argument into account, we can understand the structure of the perturbation terms of the PSA limit of the electromagnetic Casimir energy between $N$ bodies. The Casimir energy in the PSA limit has the structure of a series of infinity terms whose $n$-order term is the sum of integrals over $n+2$ bodies (which can be repeated or not) of the $(n+2)$ points local em potential. Each series term come from the expansions made in \ecref{Formula de Emig en forma de traza} and in \ecref{Born expansion of T operator} and is proportional to the product of $n+2$ permittivities. If we take the asymptotic distance limit of these integrals, we will obtain a series of $n$ points em potential now with polarizabilities instead susceptibilities as in \cite{Thiru}, but in the soft response limit. When we use perturbations of the tree term of the Born expansion, it appears divergent terms in the integrand as seen in \ecref{Divergent integrand example}. These singularities can be removed with the appropriate regularization procedure.

\section{Final Remarks}
We have calculated the full electromagnetic Casimir energy in the diluted limit between two bodies using the formalism given in \cite{Kardar-Geometrias-Arbitrarias} reobtaining the energy given in \cite{Milton} in the pure electric case. Although this formalism fails in the perfect metal case, it is valid for soft dielectric and diamagnetic bodies. This formalism can be extended for an arbitrary number of bodies. We have shown that the Casimir energy, which is a nonadittive interaction, has a superposition behavior in the first expansion of the PSA and how the integration of different $N$ point potentials appear in the perturbation series to contribute in the whole energy. Then we have a systematic procedure to calculate Casimir energies valid for soft dielectric and diamagnetic bodies.

\acknowledgements
I acknowledge helpful discussions with R.~Brito and T.~Emig.  This research was supported by projects MOSAICO, UCM/PR34/07-15859 and a FPU MEC grant.

% Apéndice
\appendix
\section*{Appendix: Obtention of the matricial Green function}
\renewcommand{\theequation}{A.\arabic{equation}}
\setcounter{equation}{0}
\label{sec:Appendix}
Here we have calculated the Casimir energy using a matricial Green function for the electromagnetic fields instead the dyadic Green function used in \cite{Kardar-Geometrias-Arbitrarias}. in this appendix we will see that both formalisms are equivalents just using as field sources the induced polarization and magnetization vectors instead the induced currents into the bodies. The induced currents into the bodies are \cite{Jackson}:
\begin{equation}
\textbf{j} = -ik\textbf{P} + \nabla\times\textbf{M}.
\end{equation}
We introduce this linear change of variable in the partition function, so there is not any new relevant term in the action of the problem. This action will be transformed in the next way:
\begin{equation}
S = \int d\textbf{x}\bar{\textbf{j}}\mathcal{G}_{0}\textbf{j},
\end{equation}
\begin{eqnarray}
S & = & \int d\textbf{x}k^{2}\bar{\textbf{P}}\mathcal{G}_{0}\textbf{P} + \int d\textbf{x}\nabla\times\bar{\textbf{M}}\mathcal{G}_{0}\nabla\times\textbf{M}\nonumber\\
  &   & + \int d\textbf{x}ik\bar{\textbf{P}}\mathcal{G}_{0}\nabla\times\textbf{M} - \int d\textbf{x}\nabla\times\bar{\textbf{M}}\mathcal{G}_{0}ik\textbf{P}.
\end{eqnarray}
Where
\begin{equation}
\mathcal{G}_{0} = \left[\delta_{ij} - \frac{1}{k^{2}}\nabla_{i}\nabla_{j}\right]G_{0},
\end{equation}
is the Green dyadic function. We integrate by parts each action term obtaining:
\begin{eqnarray}
S_{EE} & = & \int d\textbf{x}k^{2}\bar{\textbf{P}}\mathcal{G}_{0}\textbf{P},\nonumber\\
       & = & \int d\textbf{x}\bar{\textbf{P}}_{i}\left[k^{2}\delta_{ij} - \nabla_{i}\nabla_{j}\right]G_{0}\textbf{P}_{j},\nonumber\\
       & = & \int d\textbf{x}\bar{\textbf{P}}_{i}G_{ij}^{EE}\textbf{P}_{j},
\end{eqnarray}
\begin{eqnarray}
S_{EH} & = & \int d\textbf{x}ik\bar{\textbf{P}}\mathcal{G}_{0}\vec{\nabla}\times\textbf{M}, \nonumber\\
       & = & \int d\textbf{x}ik\bar{\textbf{P}}_{i}\left[\delta_{ij} - \frac{1}{k^{2}}\nabla_{i}\nabla_{j}\right]G_{0}\epsilon_{j\alpha\beta} \nabla^{\alpha}\textbf{M}^{\beta},\nonumber\\
& &
\end{eqnarray}
\begin{eqnarray}
S_{EH} & = & \int d\textbf{x}\bar{\textbf{P}}_{i}\left[ik\epsilon_{i\alpha\beta}\nabla^{\beta}\right]G_{0}\textbf{M}^{\alpha} \nonumber\\
 & &   + \int d\textbf{x}\bar{\textbf{P}}_{i}\left[ik\frac{1}{k^{2}}\nabla_{i}\nabla_{j}\epsilon_{j\alpha\beta}\nabla^{\beta}\right]G_{0}\textbf{M}^{\alpha}.
\end{eqnarray}
The second term is zero:
\begin{equation}
\epsilon_{j\alpha\beta}\nabla_{j}\textbf{M}^{\alpha}\nabla^{\beta}G_{0} = \textbf{M}\cdot\vec{\nabla}\times\vec{\nabla}G_{0} = \textbf{M}\cdot\textbf{0} = 0.
\end{equation}
So we finally get:
\begin{eqnarray}
S_{EH} & = & \int d\textbf{x}\bar{\textbf{P}}_{i}\left[-ik\epsilon_{ij\beta}\nabla^{\beta}\right]G_{0}\textbf{M}^{j},\nonumber\\
       & = & \int d\textbf{x}\bar{\textbf{P}}_{i}G_{ij}^{EH}\textbf{M}^{j}.
\end{eqnarray}
$S_{HE}$ is similar to $S_{EH}$, but with a changed sign because the complex conjugation of the induced current into the action:
\begin{equation}
S_{HE} =  - \int d\textbf{x}\nabla\times\bar{\textbf{M}}\mathcal{G}_{0}ik\textbf{P},
\end{equation}
\begin{equation}
S_{HE} = \int d\textbf{x}\bar{\textbf{M}}_{i}G_{ij}^{HE}\textbf{P}^{j} = \int d\textbf{x}\bar{\textbf{M}}_{i}\left[- G_{ij}^{EH}\right]\textbf{P}^{j}.
\end{equation}
And the last term:
\begin{equation}
S_{HH} = \int d\textbf{x}\nabla\times\bar{\textbf{M}}\mathcal{G}_{0}\vec{\nabla}\times\textbf{M},
\end{equation}
\begin{equation}
S_{HH} = \int d\textbf{x}\epsilon^{i\alpha\beta}\nabla_{\alpha}\bar{\textbf{M}}_{\beta}\left[\delta_{ij} - \frac{1}{k^{2}}\nabla_{i}\nabla_{j}'\right]G_{0}\epsilon_{jab}\nabla'^{a}\textbf{M}^{b},
\end{equation}
\begin{equation}
S_{HH} = \int d\textbf{x}\epsilon^{i\alpha\beta}\bar{\textbf{M}}_{\alpha}\nabla_{\beta}\left[\delta_{ij} - \frac{1}{k^{2}}\nabla_{i}\nabla_{j}'\right]G_{0}\epsilon_{jab}\textbf{M}^{a}\nabla'^{b},
\end{equation}
\begin{eqnarray}
S_{HH} & = & \int d\textbf{x}\bar{\textbf{M}}_{\alpha}\left(\epsilon^{i\alpha\beta}\nabla_{\beta}\delta_{ij}G_{0}\epsilon_{jab}\nabla'^{b}\right)\textbf{M}^{a} \nonumber\\
 & & - \int d\textbf{x}\bar{\textbf{M}}_{\alpha}\left(\epsilon^{i\alpha\beta}\nabla_{\beta}\frac{1}{k^{2}}\nabla_{i}\nabla'^{j}G_{0}\epsilon_{jab}\nabla'^{b}\right)\textbf{M}^{a},\nonumber\\
& &
\end{eqnarray}
Using $\epsilon^{i\alpha\beta}\delta_{i}^{j}\epsilon_{jab} = \delta_{a}^{\alpha}\delta_{b}^{\beta} - \delta_{b}^{\alpha}\delta_{a}^{\beta}$ and $\nabla_{b}\nabla'^{b}G_{0} = k^{2}G_{0}$, we get the next result:

\begin{eqnarray}
\left(\epsilon^{i\alpha\beta}\nabla_{\beta}\delta_{ij}G_{0}\epsilon_{jab}\nabla'^{b}\right) & = & \left[\delta_{a}^{\alpha}\delta_{b}^{\beta} - \delta_{b}^{\alpha}\delta_{a}^{\beta}\right]\nabla_{\beta}\nabla'^{b}G_{0}, \nonumber\\
 & = & \left[\delta_{a}^{\alpha}\nabla_{b}\nabla'^{b} - \nabla_{a}\nabla'^{\alpha}\right]G_{0},\nonumber\\
 & = & \left[\delta_{a}^{\alpha}k^{2} - \nabla_{a}\nabla'^{\alpha}\right]G_{0}.
\end{eqnarray}
The other term requires even a more tedious work, but it is easy to obtain that

\begin{equation}
\frac{-1}{k^{2}}\epsilon^{i\alpha\beta}\epsilon_{jab}\nabla_{\beta}\nabla_{i}\nabla'^{j}\nabla'^{b}G_{0} = 0,
\end{equation}
because this differential operator is zero. So $S_{HH}$ is
\begin{eqnarray}
S_{HH} & = & \int d\textbf{x}\bar{\textbf{M}}_{\alpha}\left(\epsilon^{i\alpha\beta}\nabla_{\beta}\delta_{ij}G_{0}\epsilon_{jab}\nabla'^{b}\right)\textbf{M}^{a}, \nonumber\\
 & = & \int d\textbf{x}\bar{\textbf{M}}_{\alpha}\left[\delta_{a}^{\alpha}k^{2} - \nabla_{a}\nabla'^{\alpha}\right]G_{0}\textbf{M}^{a},\nonumber\\
 & = & \int d\textbf{x}\bar{\textbf{M}}_{i}G_{ij}^{HH}\textbf{M}_{j}.
\end{eqnarray}
After a Wick rotation, we obtain the used form of the matricial Green function, which components are, using $G_{0}(R,k) = \frac{e^{-kR}}{4\pi R}$ and $R = \abs{\textbf{r} - \textbf{r}'}$:
\begin{eqnarray}
G_{0ij}^{EE}(R,k) & = & \left[k^{2}\delta_{ij} + \nabla_{i}\nabla_{j}'\right]G_{0}(R,k),\label{Acoplo EE funcion Green}\\
G_{0ij}^{EH}(R,k) & = & - k\epsilon_{ijk}\nabla_{k}G_{0}(R,k),\label{Acoplo EH funcion Green}\\
G_{0ij}^{HE}(R,k) & = & k\epsilon_{ijk}\nabla_{k}G_{0}(R,k),\label{Acoplo HE funcion Green}\\
G_{0ij}^{HH}(R,k) & = & \left[k^{2}\delta_{ij} + \nabla_{i}\nabla_{j}'\right]G_{0}(R,k)\label{Acoplo HH funcion Green}.
\end{eqnarray}
Which are the results used in \ecref{Formula de Emig en soft limit}. It is also possible to obtain the same result from fluctuation - dissipation theorem, as made in \cite{Agarwal}.

%Para Bibtex, se usan las siguientes instrucciones:
%\bibliography{References}
%\bibliographystyle{unsrt}
% Para el caso que nos ocupa, lo haré de la forma cutre:

\end{document}